# Versatile van der Waals Heterostructures of γ-GeSe with h-BN/Graphene/MoS$_2$

Changmeng Huan[1,2], Pu Wang[1,2], Bingtao Liu[1,2], Binghan He[1,2], Yongqing Cai[3]*, Qingqing Ke[1,2]*

[1]School of Microelectronics Science and Technology, Sun Yat-sen university, Zhuhai 519082, China

[2]Guangdong Provincial Key Laboratory of Optoelectronic Information Processing Chips and Systems, Sun Yat-sen University, Zhuhai 519082, China

[3]Joint Key Laboratory of the Ministry of Education, Institute of Applied Physics and Materials Engineering, University of Macau, Taipa, Macau, China

\* Corresponding authors

E-mail: yongqingcai@um.edu.mo; keqingq@mail.sysu.edu.cn

**Abstract:**

Recent discovery of a novel hexagonal phase of GeSe (γ-GeSe) has triggered great interests in nanoelectronics applications owing to its electrical conductivity of bulk phase even higher than graphite while its monolayer is a semiconductor. For potential applications, construction of functional two-dimensional (2D) contacts is indispensable. Herein, via first-principles calculations, we propose the design of van der Waals heterostructures (vdWHs) of γ-GeSe contacting respectively with graphene, 2D h-BN and MoS$_2$, as representatives of metallic, insulator, and semiconductor partners. Our work shows that the h-BN or graphene layer donates electrons to the γ-GeSe layer, resulting in *n* doping in γ-GeSe, while the MoS$_2$ layer accepts electrons from the γ-GeSe layer leading to *p* doping of the latter. The γ-GeSe/BN heterostructure has a type-I band alignment with large band offsets, indicating that BN can be used as an effective passivating layer to protect γ-GeSe from its environmental disturbance while maintaining its major electronic and optical characteristics. For γ-GeSe/graphene heterostructure, it is prone to have a very low‐Schottky barrier down to tens of meV, easily overcome by thermal excitation, which can be tunable by strain and external electric field. The γ-GeSe/MoS$_2$ vdWH forms a Z-scheme interface, which is beneficial for carriers splitting and photon utilization. Our work indicates that γ-GeSe can be well passivated by BN, and form intimate contact with graphene for high charge injection efficiency and with MoS$_2$ for efficient carriers splitting for redox reactions.

# 1. Introduction

The successful exfoliation of graphene initiated a new era of two-dimensional (2D) materials.[1] A large family of 2D materials including transition metal dichalcogenides (TMDs),[2,3] phosphorene,[4,5] hexagonal boron nitride (h-BN),[6,7] MXene,[8,9] have attracted extensive attention owing to their distinct mechanical, electronic, and optical properties. A single type of a 2D material generally cannot meet the requirements of practical applications, and construction of van der Waals heterostructures (vdWHs) with integrating different types of 2D layers together renders additional functionality and promoted performance.[10-15] The vdWHs also lead to exotic electronic properties owing to the commensurate lattice or misfit stacking allowing manipulation of quasi-particles dynamics within the vdW gap. In particular, interlayer excitons in vdWHs, providing optically addressable spin and valley degrees of freedom and long lifetimes,[16] have become the hottest research area today. The exciton transistors,[17] exciton router,[18] moiré excitons,[19, 20] and the photo- and electro-luminescence from interlayer excitons[21, 22] have been achieved in vdWHs.

As one member of group IV monochalcogenide MX (M = Ge, Sn, Pb; X = S, Se, Te), the layered α-GeSe has been extensively studied owing to its rich polymorphs, phosphorene-like structure, excellent stability and environmental sustainability.[23-26] Recently, a novel hexagonal phase of GeSe (γ-GeSe) with two merged blue-phosphorus-like structure, has been synthesized for the first time.[27] This new phase shows striking features with a graphene-like semimetallic behavior, albeit highly dependent on thickness, and its bulk phase shows a surprisingly better metallic conductivity than graphite.[27] Further, a strain-tunable spontaneous polarization and spin-splitting in 2D form has been predicted through theoretical calculations.[28] Consisting of solely s- and p-orbitals, the γ-GeSe shows the intriguing property of a superior conduction in bulk form while a semiconducting nature in monolayer. This renders it as an ideal platform as a passive layer, through forming heterostructures with other 2D materials, for rectifying resistivity or contacts in 2D nanoelectronics. However, how does such vdW hetero-interface consisting of γ-GeSe form, and its related energetics and electronic properties such as lattice registration and band alignment are still unknown albeit critically important.

In this work, we attempt to design the γ-GeSe based 2D heterostructures for various functional contacts via first-principles calculations. We narrowed down its potential partners to 2D graphene, BN, and MoS$_2$ as representatives of metallic, insulator, and semiconducting contacts, respectively. The vertically stacked γ-GeSe/BN, γ-GeSe/graphene, and γ-GeSe/MoS$_2$ vdWHs largely maintain the electronic structure of GeSe due to its weak van der Waals force. The γ-GeSe/BN and γ-GeSe/MoS$_2$ show a Type-I and type-II band alignment (Z-scheme interface), respectively, while the γ-GeSe/graphene possesses efficient thermally activated contact with a low Schottky barrier of tens of meV. Our findings will be useful for the effective passivation of γ-GeSe and functional integration for 2D nanoelectronics devices.

## 2. Computational methods

The first-principles calculations were carried out by the Vienna ab-initio simulation package (VASP) with projector augmented wave (PAW) method.[29, 30] The generalized gradient approximation (GGA) with the Perdew−Burke−Ernzerhof (PBE) functional[31] was employed to describe the exchange-correlation energy. A plane-wave basis set with a cutoff of 400 eV was used for all calculations. The Brillouin zones were integrated using the Gamma-centered k-points sampling with a reciprocal space resolution higher than $2\pi \times 0.02$ Å$^{-1}$ for geometry optimization and electronic structure calculations. The convergence thresholds for residual force and total energy were 0.005 eV/Å and $10^{-8}$ eV, respectively. A vacuum layer over 20 Å in *z*-direction was set to avoid the periodic interaction. The DFT-D3 corrections[32, 33] was used for van der Waals interaction. The Heyd-Scuseria-Ernzerhof (HSE06) hybrid functional[34] was also used to evaluate the electronic structures. The thermal stability of the vdWHs was examined by the ab-initio molecular dynamics (AIMD) simulations at 300 K for 10 ps with a time step of 2 fs.[35]

The interfacial binding energy ($E_b$) of the vdWHs was calculated by[36]

$$E_b = (E_{GeSe/X} - E_{GeSe} - E_X)/S \qquad (1)$$

where $E_{GeSe/X}$, $E_{GeSe}$, and $E_X$ represent the total energy of the GeSe/*X* (*X* = h-BN, graphene, 2H-MoS$_2$) vdWH, the GeSe layer, and the *X* layer, respectively, and *S* represents the surface area.

The differential charge density (DCD) $\Delta\rho(r)$ is defined as

$$\Delta\rho = \rho_{GeSe/X} - \rho_{GeSe} - \rho_X \qquad (2)$$

where $\rho_{GeSe/X}$, $\rho_{GeSe}$, $\rho_X$ are the charge densities of the GeSe/*X* vdWH, the GeSe layer, and the *X* layer, respectively.

## 3. Results and Discussion

**Geometric and electronic structure of the freestanding monolayers:**

Before studying the GeSe/X vdWHs, the freestanding γ-GeSe and X monolayers were first investigated. We calculated their band structures by the PBE (Fig.1) and HSE06 (Fig. S1) functionals, the results are summarized in Table 1. The lattice constants of γ-GeSe, h-BN, graphene, and MoS$_2$ are 3.76, 2.51, 2.47, and 3.15 Å, respectively, agreeing well with the results reported previously.[27, 37-39] The γ-GeSe monolayer exhibits an indirect bandgap of 0.60 eV in PBE method, where the conduction band minimum (CBM) is located at the Γ point and the valence band maximum (VBM) is located at the $K_1$ point between the Γ and K points. The h-BN monolayer shows a wide bandgap of 4.65 eV in PBE method, behaving as an insulator. Graphene acts as the well-known half-metal with Dirac-cone at the K point, while the single-layer $MoS_2$ possesses a direct bandgap of 1.77 eV at the K point. Importantly, for γ-GeSe, there are additional satellite states at the $M_1$ and M points in the top valence band and bottom conduction band of γ-GeSe, respectively, where the $M_1$ point is only ~0.03 eV lower than the VBM and the M point only ~0.10 eV higher than the CBM, suggesting high chance for thermal induced excitations. Compared with other 2D materials, the presence of an indirect band gap and hole pockets states in γ-GeSe will be highly appealing for thermoelectric with promoted density of carriers at band edges and for solar cell applications due to suppressed direct electron-hole recombination at Γ.

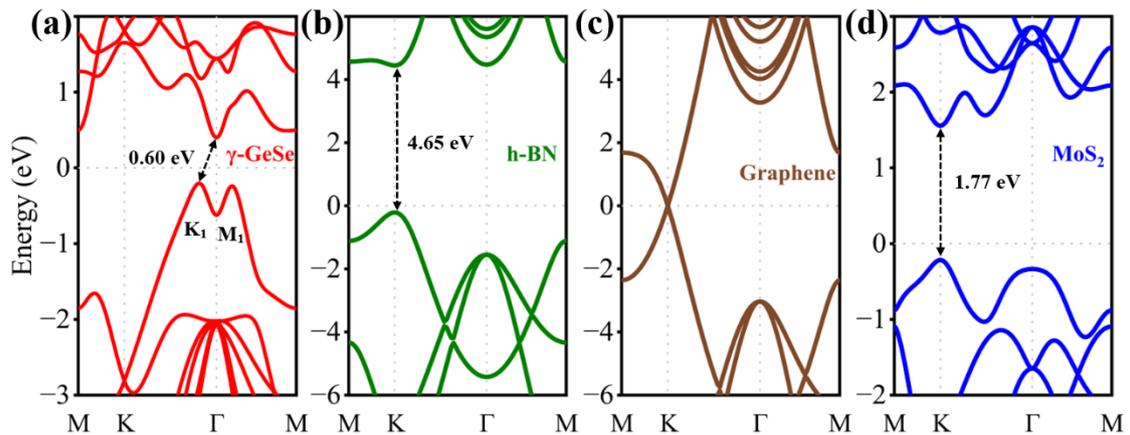

**Fig. 1** The band structures of γ-GeSe (a), h-BN (b), graphene (c), and 2H-$MoS_2$ (d) monolayers by using PBE method. The Fermi level is aligned to 0 eV.

**Table 1.** The equilibrium lattice constant, bandgap ($E_g$), and work function ($\varphi$) of the monolayer materials. $E_g$ was obtained by both PBE and HSE06 functionals.

| Monolayer | Lattice constant (Å) | PBE $E_g$ (eV) | HSE06 $E_g$ (eV) | PBE $\varphi$ (eV) |
|:---:|:---:|:---:|:---:|:---:|
| γ-GeSe | 3.76 | 0.60 | 0.99 | 4.47 |
| h-BN | 2.51 | 4.65 | 5.69 | 5.60 |
| Graphene | 2.47 | / | / | 4.29 |
| MoS$_2$ | 3.15 | 1.77 | 2.34 | 5.69 |

**Geometric structure and stability of the γ-GeSe/*X* vdWHs:**

To reduce the lattice mismatch, we built the γ-GeSe/BN and γ-GeSe/graphene vdWHs using 2×2×1 γ-GeSe, 3×3×1 h-BN and graphene supercells. The γ-GeSe/MoS$_2$ vdWH is composed of √3×√3×1 γ-GeSe supercell and 2×2×1 MoS$_2$ supercell. Possible stable stacking order and configurations are considered in Fig. S2 and S3 of the supporting information. It can be found that relative shift and different stacking order at the interface have little effect on the γ-GeSe/BN and γ-GeSe/graphene vdWHs, while the configuration-1 of γ-GeSe/MoS$_2$ vdWH with γ-GeSe layer at the bottom shows a lower energy and is thus more stable. The lattice structures and lattice mismatches of the vdWHs with the most stable stacking order and configuration are shown in Fig. 2 and Table S1, respectively. The lattice constants of γ-GeSe/BN, γ-GeSe/graphene, and γ-GeSe/MoS$_2$ vdWH are 7.53, 7.42, and 6.39 Å with the lattice mismatches of 0.13%, 1.35%, and 1.92%, respectively, which are within a reasonable range.

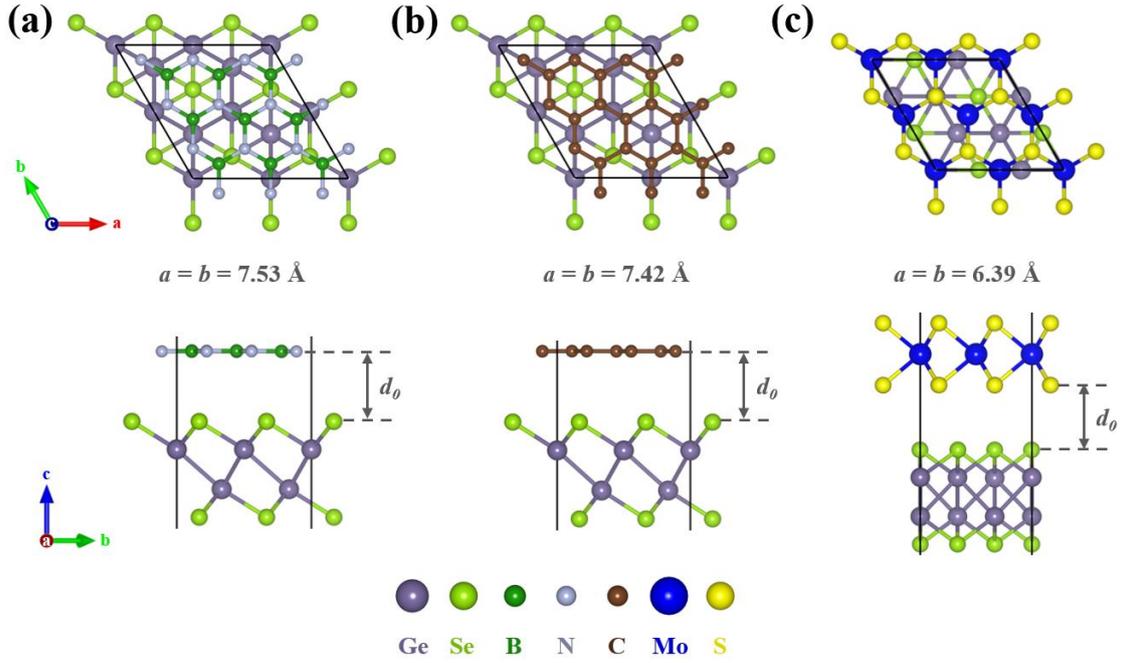

**Fig. 2** Construction of commensurate heterostructures of γ-GeSe/BN (a), γ-GeSe/graphene (b), and γ-GeSe/MoS$_2$ (c) with the top and front views. The black lines show the cell of the bilayer model with the good lattice match.

The $E_b$ was calculated to evaluate the relative stability of the vdWHs, and the results of $E_b$ as a function of interlayer distances ($d_0$) are plotted in Fig. 3. It can be found that all three vdWHs show negative $E_b$ values, indicating that they are all thermodynamically favored. The optimal interlayer distances are 3.41, 3.46, and 3.25 Å, with the minimum $E_b$ values being −15.27, −16.29, and −20.90 meV/Å$^2$ for GeSe/BN, GeSe/graphene, and GeSe/MoS$_2$ vdWHs, respectively. For all the three vdWHs, we found that the $d_0$ and $E_b$ values are in the same range as other vdWHs, such as MoX$_2$/WX$_2$ (X = S, Se) and MoS$_2$/graphene.[40-43] Furthermore, AIMD simulations at 300 K were conducted further to verify the thermostability of γ-GeSe/*X* vdWHs, as shown in Fig. S4. The total energy undergoes small fluctuation, and the structural snapshots at 10 ps show that the atoms are only slightly displaced near their equilibrium positions with no structural corruption, confirming the excellent thermostability.

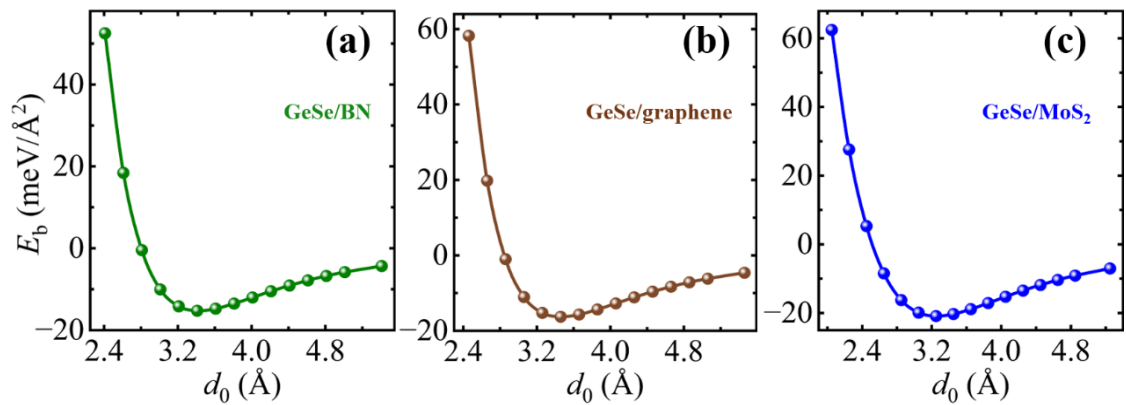

**Fig. 3** The variation of $E_b$ as a function of $d_0$ in γ-GeSe/BN (a), γ-GeSe/graphene (b), and γ-GeSe/MoS$_2$ (c) vdWHs.

**Electronic Properties of the GeSe/*X* vdWHs:**

Considering that lattice mismatch can impose strain ($\varepsilon$) on the constituent layers and change the electronic structure of both layers, impeding drawing conclusions about interlayer interactions in vdWHs. To decouple the strain effects and interlayer interactions, the band structure of each monolayer supercell fixing to the lattice constant of vdWHs was first calculated as a reference. Compared to the unit cell, the shape of band structure of the supercell may change due to band folding but keep the bandgap value (i.e., the energy positions of the CBM and VBM remain unchanged). For γ-GeSe/BN (Fig. S5a and 5d), the band gaps of γ-GeSe ($\varepsilon = +0.13\%$) and h-BN ($\varepsilon = 0$) are preserved due to the small intentional created strains arising from lattice mismatch, while the direct bandgap of h-BN in supercell is folded to the Γ point instead of the K point in the unit cell. In the case of γ-GeSe/graphene (Fig. S5b and 5e), the indirect bandgap of γ-GeSe monolayer is preserved, while the compressive strain ($\varepsilon = -1.33\%$) induced by lattice mismatch makes a transition of the CBM from the Γ point to the point between Γ and M points, leaving a bandgap of 0.59 eV. The small tensile strain ($\varepsilon = +0.13\%$) has a little effect on the electronic properties of graphene, and the Dirac-cone is folded to the Γ point instead of the K point in the unit cell as the graphene layer is a supercell.[42] In the case of γ-GeSe/MoS$_2$ (Fig. S5c and 5f), the VBM of γ-GeSe monolayer ($\varepsilon = -1.88\%$) is shifted from the initial $K_1$ point to the $M_1$ point, and the CBM moves to the point between the K and M points, resulting in a bandgap of 0.54 eV. Similarly, the VBM of MoS$_2$ ($\varepsilon = +1.43\%$) moves from the K point to the Γ point, while the CBM remains at the K point, resulting in an indirect band gap of 1.56 eV.

The projected band structures of the three vdWHs are shown in Fig. 4. For the γ-GeSe/BN vdWH in Fig. 4a, the bandgap (0.60 eV), the CBM and VBM match well with those of γ-GeSe monolayer, implying that the electronic states of GeSe are energetically well buried between those of BN. There are obvious hybridizations between the γ-GeSe and h-BN in the valence and conduction bands due to the interlayer interactions. The plot of charge densities (bottom panel) indicates that the electrons at both the CBM and VBM are entirely localized in the γ-GeSe layer. This suggests that the BN makes negligible effect on the electronic properties of γ-GeSe and could be a good passivation layer to cover and protect the underneath γ-GeSe layer.

For the γ-GeSe/graphene vdWH in Fig. 4b, there are several interesting features in the band structures. First, the band structures of this vdWH seem to be a direct combination of band structures of γ-GeSe (Fig. S5b) and graphene (Fig. S5e) monolayers. This gives rise to the preservation of intrinsic Dirac-cone of graphene and the indirect bandgap of γ-GeSe. Second, there is a small bandgap opening (10 meV) of graphene, smaller than the thermal fluctuation at room temperature (26 meV). Similar phenomenon occurred in the graphene/BiI$_3$ (17 meV) and CsPbBr$_3$/graphene (29 meV) heterostructures,[44, 45] which is due to the breaking of the structural symmetry and the asymmetric potential.[46] Furthermore, the resonant valence and conduction bands show an obvious hybridization between the γ-GeSe and graphene, reflecting the interlayer interactions. The charge distributions (bottom panel) show that the electrons at both the CBM and VBM are entirely localized in the graphene layer. The Fermi level crosses graphene but close to the valence top of γ-GeSe. Therefore, the γ-GeSe/graphene vdWH belongs to a metal-semiconductor (MS) contact.

For the γ-GeSe/MoS$_2$ vdWH (Fig. 4c), it exhibits an indirect band gap (0.10 eV) with the CBM and VBM located at the K point of the MoS$_2$ side and the M$_1$ point (between the Γ and M points) of the γ-GeSe side, respectively. While the vdWH retains the basic features of the band structures of γ-GeSe (Fig. S5c) and MoS$_2$ (Fig. S5f), the valence and conduction bands show strong hybridization between γ-GeSe and MoS$_2$, and the VBM of MoS$_2$ is shifted from the Γ (Fig. S5f) to the point between the Γ and M points (Fig. 4c), indicating the interlayer interactions. The charge distributions in the bottom panel of Fig. 4c also indicates that the CBM states reside in the MoS$_2$ layer, while the VBM states are located in the γ-GeSe layer. This spatially separated VBM and CBM facilitates the separation of electron and holes which is highly appealing for photocatalysis and solar cell applications.

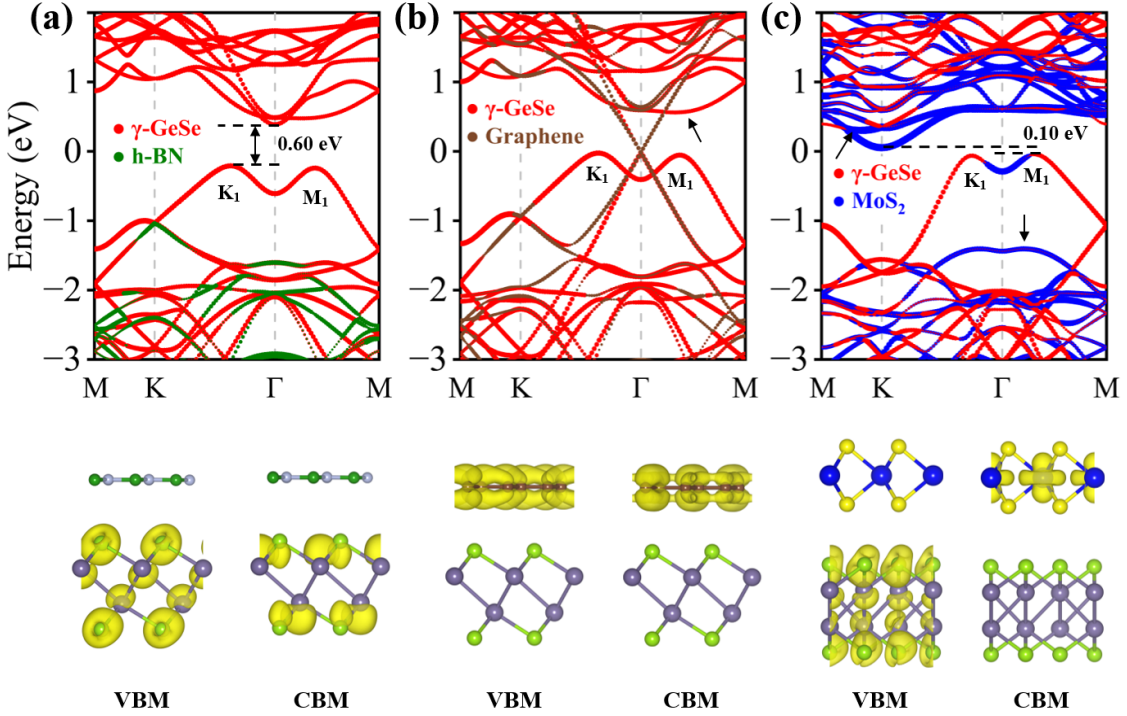

**Fig. 4** Projected band structures (top panel) and the partial charge densities at the CBM and VBM (bottom panel) of γ-GeSe/BN (a), γ-GeSe/graphene (b), and γ-GeSe/MoS$_2$ (c) vdWHs. The Fermi level is set to 0.

**Charge redistribution and transfer:**

To further reveal the interlayer interactions in these GeSe based vdWHs, the interfacial charge transfer analyses are given. The DCD Δρ(z) along the normal direction (z) to the surface, which is calculated by integrating Δρ(r) within the basal plane at the z point. The amount of transferred charge ΔQ(z) at z point is given by[47, 48]

$$\Delta Q(z) = \int_{-\infty}^{z} \Delta \rho(z') \, dz' \quad (3)$$

The total amount of transferred charge is determined by the value of ΔQ(z) at lobe in the interface, where the Δρ(z) curve shows a zero value. The results of Δρ(z) and ΔQ(z) as a function of the position (z) are plotted in Fig. 5(a-c). It can be seen that the h-BN or graphene layers donate electrons to the γ-GeSe layer, resulting in *n*-doping in γ-GeSe and *p*-doping in h-BN or graphene, while the $MoS_2$ layer accepts electrons from the γ-GeSe layer. The isosurfaces of the DCD are plotted in the bottom panels, where the depletion and accumulation of electrons across the interface are vividly depicted. The total electrons gain of γ-GeSe in GeSe/BN and GeSe/graphene are $6.52\times10^{-4}$ e/Å$^2$ and $3.77\times10^{-4}$ e/Å$^2$, respectively, while the total electrons loss of γ-GeSe in GeSe/$MoS_2$ is $5.09\times10^{-4}$ e/Å$^2$, which can lead to the formation of interfacial dipoles.

Fig. 6 shows the plane-averaged electrostatic potential along the *z*-direction of γ-GeSe/BN, γ-GeSe/graphene and γ-GeSe/$MoS_2$. It can be found that there is difference in work function on both sides of the γ-GeSe/*X* vdWHs. This is due to the interfacial dipoles caused by interlayer charge transfer, as shown in Fig. 5, where electrons flow from the h-BN or graphene (γ-GeSe) layer to the γ-GeSe ($MoS_2$) layer. The interfacial dipoles can be defined via a potential step $\Delta V$ at the interface. The potential step can be defined as $\Delta V = \varphi_{GeSe} - \varphi_X$, where $\varphi_{GeSe}$ and $\varphi_X$ are the work functions of γ-GeSe and *X* sides in the γ-GeSe/*X* vdWHs, respectively. The calculated $\Delta V$ in γ-GeSe/BN, γ-GeSe/graphene and γ-GeSe/$MoS_2$ vdWHs is 0.20, 0.12 and 0.16 eV, respectively. Furthermore, the potential drop ($\Delta V_d$) across the bilayers is found to be 9.35, 12.14 and 6.40 eV, respectively. Such large potential difference implies a strong electrostatic field across the interface, which may significantly influence the carrier dynamics and charge injection.

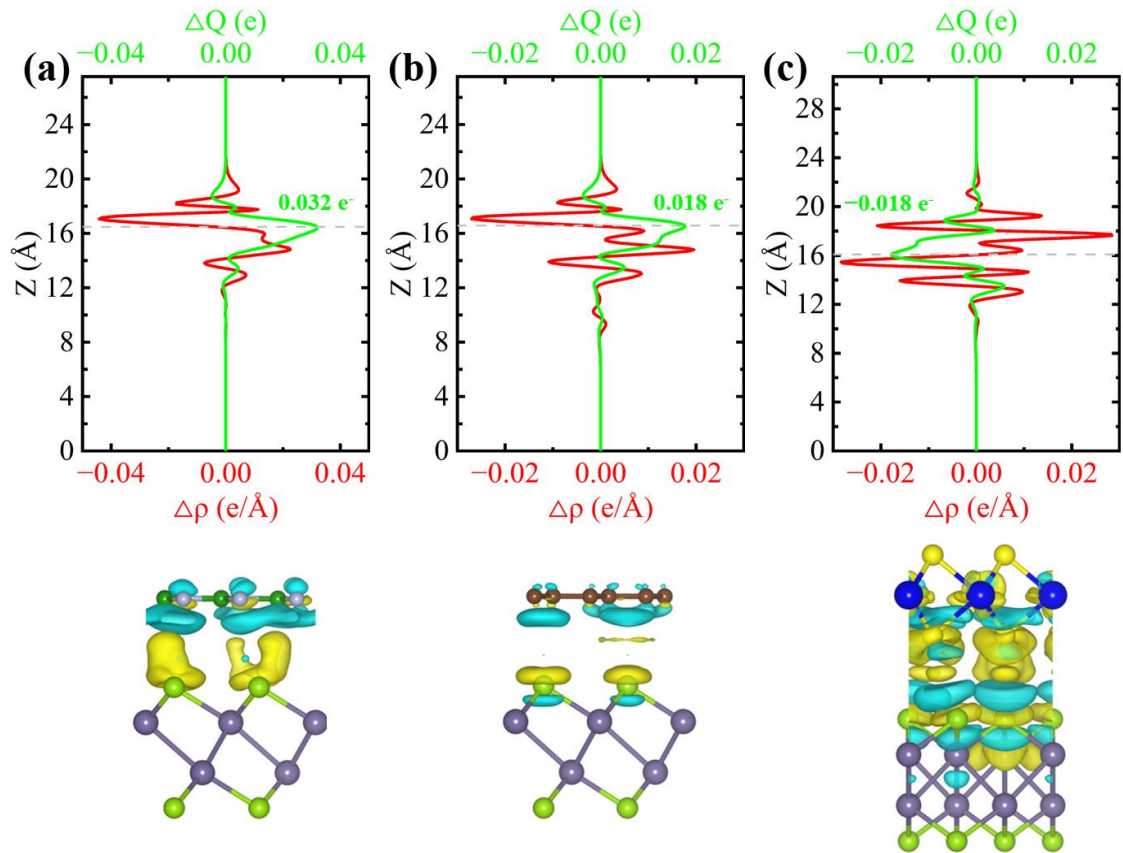

**Fig. 5** Plane-averaged DCD Δ$ρ(z)$ and the amount of transferred charge along the *z*-direction Δ$Q(z)$ for γ-GeSe/BN (a), γ-GeSe/graphene (b) and γ-GeSe/MoS$_2$ (c). Bottom panels: front view of the DCD isosurfaces with the isovalue of $10^{-4}$ e/Å$^3$. The yellow (blue) color denotes accumulation (depletion) of electrons.

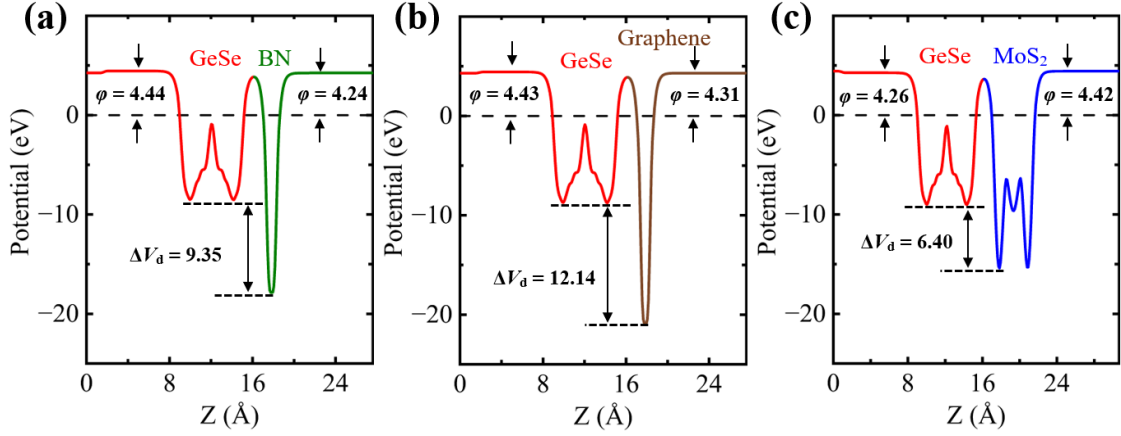

**Fig. 6** The calculated electrostatic potential along the *z*-direction for γ-GeSe/BN (a), γ-GeSe/graphene (b) and γ-GeSe/MoS$_2$ (c) vdWHs. The Fermi level is set to 0.

**Band alignment of the γ-GeSe/*X* vdWHs:**

The band alignment is one of the most important properties for heterostructures which determines the flow of charge carriers across the interface and the types of contacts, i.e Ohmic or Schottky types. The band alignments of γ-GeSe/*X* vdWHs are calculated and shown in Fig. 7. It can be found that the band alignments of GeSe/BN and GeSe/MoS$_2$ indicate the straddling gap (Type-I) and staggered gap (type-II), respectively, while the GeSe/graphene belongs to a metal-semiconductor contact. The conduction (valence) band offsets, derived by the relative position difference of CBM (VBM) at both sides of the heterostructure, are 2.56 (1.38) and 0.25 (1.36) eV for GeSe/BN and GeSe/MoS$_2$, respectively. For the GeSe/BN vdWH, the band offsets are so large that h-BN is intrinsically transparent with respect to γ-GeSe. Furthermore, the electronic structure of γ-GeSe monolayer is preserved after contacting the h-BN layer, indicating that h-BN is stable in contact with γ-GeSe and can be a good covering or protecting layer to prevent environmental molecules from γ-GeSe and enhance its stability.

For γ-GeSe/MoS$_2$ vdWH, it is a type-II band alignment semiconductor, allowing a fast separation of photo-excited carriers. The calculated optical absorption of γ-GeSe/MoS$_2$ vdWH and corresponding monolayers are shown in Fig. 8a. Compared with the free-standing monolayers, the optical absorption of vdWH presents an obvious enhancement, implying a high efficiency of solar energy utilization. This may be due to the resonant electronic states caused by the charge transfer and interlayer coupling effects.[40] Strong optical absorption tends to generate a large number of photo-generated carriers, it is of great significance to analyze the carrier transfer. Generally, there are two paths for carrier transfer (Fig. 8b): Path-1 refers to the interlayer carriers recombination, in which the holes on γ-GeSe side combine with the electrons on MoS$_2$ side; Path-2 is interlayer carrier transfer, where electrons (holes) transfer from the CBM (VBM) of γ-GeSe (MoS$_2$) to the CBM (VBM) of MoS$_2$ (γ-GeSe). The two paths occur simultaneously and competes with each other.[49] Because the work function of γ-GeSe monolayer (4.47 eV) is lower than that of MoS$_2$ monolayer (5.69 eV), the electrons on γ-GeSe side will flow into MoS$_2$ side when the vdWH is formed, resulting in a built-in electric field pointed from γ-GeSe to MoS$_2$ and introducing an upward (downward) band bending in γ-GeSe (MoS$_2$) side, as shown in Fig. 8b. Path-1 is promoted by the built-in electric field while Path-2 is suppressed by band bending, making the GeSe/MoS$_2$ vdWH an Z-scheme interface instead of the traditional type II interface.[50, 51] In traditional type-II interface (as shown in Fig. S6), electrons and holes are separated by an internal electric field but are bound by Coulomb interaction to form interlayer excitons, promising for excitonic devices.[52, 53] However, in the Z-scheme interface, the internal electric field and Coulomb interaction together drive interlayer excitons recombination along path-1. Therefore, the photo-excited intralayer excitons have a unidirectional separation, and meanwhile the electrons (holes) in MoS$_2$ (γ-GeSe) are subjected to interlayer annihilation following path-1. The carriers clustered in the VBM of MoS$_2$ and the CBM of γ-GeSe will retain a high potential, which is beneficial for optoelectronic devices and solar energy applications.[54]

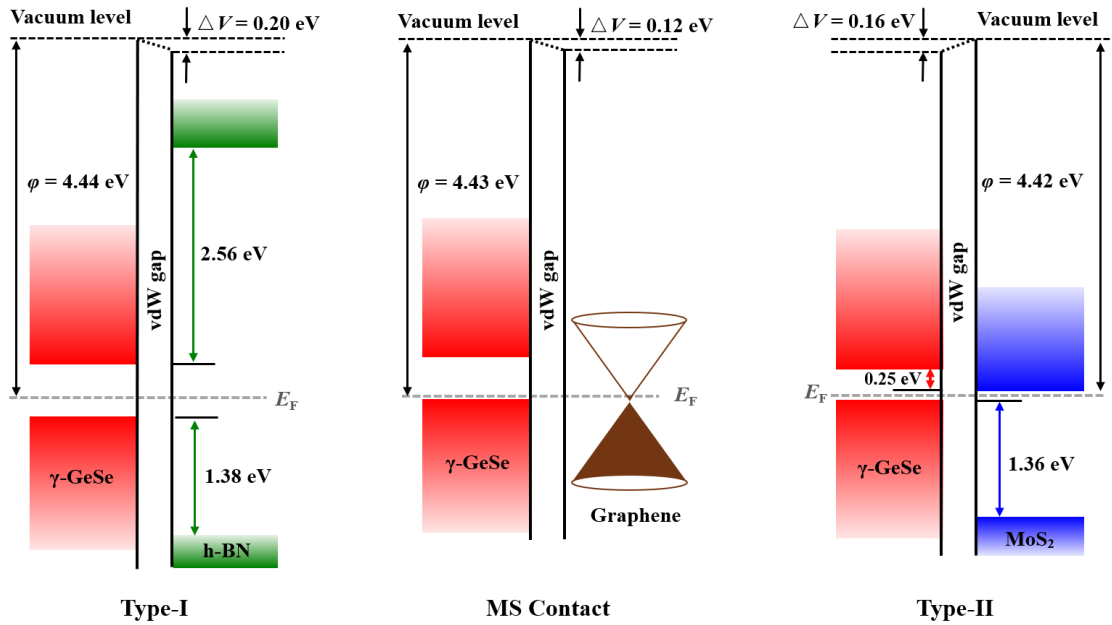

**Fig. 7** The band alignment diagrams of γ-GeSe/BN, γ-GeSe/graphene and γ-GeSe/MoS$_2$ vdWHs.

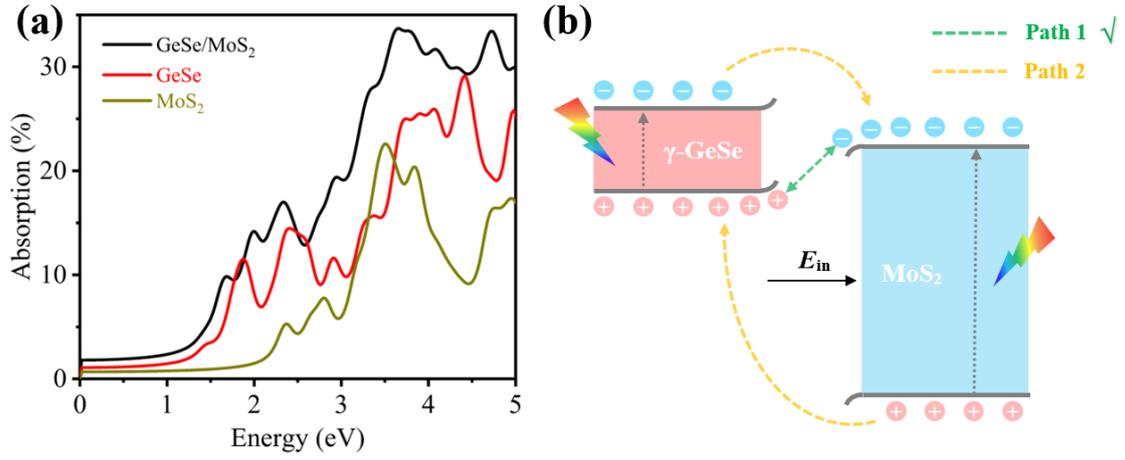

**Fig. 8** (a) The absorption spectrum of freestanding γ-GeSe and MoS$_2$ monolayers and the γ-GeSe/MoS$_2$ vdWH with HSE06 functionals. (b) The photo-induced charge transfer path of the γ-GeSe/MoS$_2$ vdWH.

As for the γ-GeSe/graphene, depending on the position of the Fermi level relative to the VBM and CBM of the semiconducting γ-GeSe monolayer, we can determine whether it forms a Schottky or an Ohmic contact. According to the Schottky-Mott rule, the *n*-type (*n*-SC) and *p*-type Schottky contact (*p*-SC) can be associated with an activation barrier as[55]

$$\Phi_n = E_{CBM} - E_F \qquad (4)$$

$$\Phi_p = E_F - E_{VBM} \qquad (5)$$

where $\Phi_n$ and $\Phi_p$ represent the *n*-type and *p*-type Schottky barrier of the heterostructure. $E_{CBM}$, $E_{VBM}$, and $E_F$ stand for the energy positions of the CBM, VBM, and Fermi level. The obtained $\Phi_n$ and $\Phi_p$ of the heterostructure are 0.56 and 0.03 eV, respectively, implying that it forms *p*-SC with an extremely small barrier of 0.03 eV. Meaningfully, we found that the Schottky barrier of γ-GeSe/graphene vdWH is strain and electric-field tunable. As shown in Fig. 9a and b, by applying compressive (−) and tensile (+) strains to the vdWH, the *p*-SC (without strain) can transform to *n*-SC ($\varepsilon = +6\%$, $\Phi_n$ = 0.16 eV < $\Phi_p$ = 0.25 eV) or to Ohmic contact ($\varepsilon = -4\%$, $\Phi_p$ = 0). The strains can alter the positions of band edges and bandgap size of γ-GeSe and slightly open the Dirac-cone (2 to 36 meV) of graphene without breaking the structure (Fig. S7). Furthermore, the strains also modulate interfacial charge transfer (Fig. S8), where compressive strain can decrease the charge transfer. Those together results in the strain-tunable Schottky barrier of γ-GeSe/graphene vdWH.

On the other hand, the Schottky barrier can be effectively tuned by external electric field ($E_{ex}$). The Schottky barrier and contact types of γ-GeSe/graphene vdWH as a function of $E_{ex}$ are plotted in Fig. 9c, where $E_{ex}$ is positive in the direction from graphene to γ-GeSe and vice versa. It can be found that the negative $E_{ex}$ results in an increase in $\Phi_p$ and a decrease in $\Phi_n$. The γ-GeSe/graphene vdWH maintains *p*-SC ($E_{ex}$ < −0.32 eV/Å), while it transits into *n*-SC ($E_{ex}$ > −0.32 eV/Å). The positive $E_{ex}$ reduces $\Phi_p$ to approximately zero (< 0.01 eV), indicating that $E_{ex}$ can convert γ-GeSe/graphene vdWH from *p*-SC to p-type Ohmic contact (*p*-OC). The above results indicate that the γ-GeSe/graphene vdWH can be used as Ohmic/Schottky contacts for high-efficiency nanodevices.

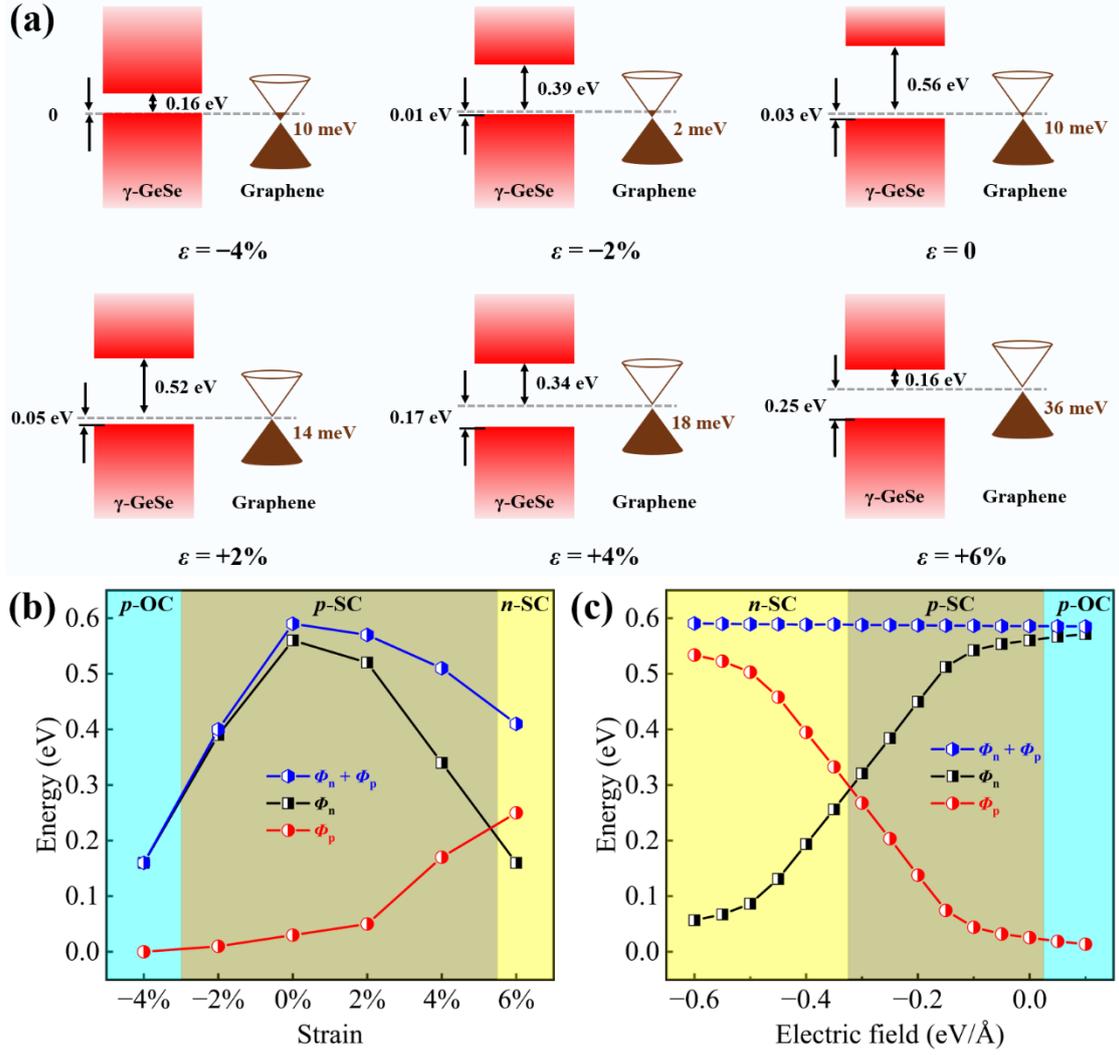

**Fig. 9** (a) Strain-tunable band edges and Schottky barrier of γ-GeSe/graphene vdWH referring to the Fermi level (dash line). The Schottky barrier of γ-GeSe/graphene vdWH as a function of strains (b) and external electric fields (c).

## 4. Conclusions

In summary, based on first-principles calculations, we examined the recently synthesized novel γ-GeSe material with respect to its potential vdWHs integrating with BN, graphene and MoS$_2$. Our results show that vdW forces between the layers largely maintain the electronic properties of respective components albeit with interfacial charge transfer. For the γ-GeSe/BN heterostructure, the h-BN layer is intrinsically transparent with respect to γ-GeSe which is highly desired for keeping its optoelectronic property and allow it be used as an effective capping layer to maintain structural stability. For the γ-GeSe/graphene semiconductor-metallic contact, it tends to have a very small Schottky barrier allowing thermally excited injection of carriers. The GeSe/MoS$_2$ vdWH forms a Z-scheme interface with suggesting a high efficiency of solar energy utilization, which is promising for optoelectronic devices. Our work would set as the theoretical foundation for the potential fabrication of γ-GeSe based 2D hybrid structures and devices.

## Conflicts of interest

The authors declare no competing financial interest.

## Acknowledgments

The authors acknowledge the funding support from the 100 Talents Program of Sun Yat-sen University (Grant 76220-18841201), Guangdong Basic and Applied Basic Research Foundation (No. 2021B1515120025), the Natural Science Foundation of China (Grant 22022309), and the Natural Science Foundation of Guangdong Province, China (2021A1515010024), the University of Macau (SRG2019-00179-IAPME, MYRG2020-00075-IAPME), the Science and Technology Development Fund from Macau SAR (FDCT-0163/2019/A3).